\documentclass[a4paper,amsmath,amssymb]{article}
\usepackage{graphicx}
\usepackage{dcolumn}
\usepackage{bm}

\title{Kink propagation in a two-dimensional curved Josephson junction}
\author{C. Gorria$^{1,2}$, Yu.~B. Gaididei$^{3}$, M.~P. Soerensen$^{1}$, 
P.~L. Christiansen$^{1}$, and J.~G. Caputo$^{4}$ \\
$^1$Section of Mathematical Physics, IMM, Technical
University \\ of Denmark, DK-2800, Kgs. Lyngby, Denmark, \\
$^2$Dept. of Applied Mathematics and Statistics, University 
\\ of the Basque Country, E - 48080 Bilbao, Spain, \\
$^3$Bogolyubov Institute for Theoretical Physics, 252143 Kiev, Ukraine, \\
$^4$Laboratoire de Mathematiques, INSA de Rouen, B.P. 08,
76131 \\ Mont-Saint-Aignan Cedex, France
and Laboratoire de Physique theorique \\ et modelisation,
Universite de Cergy-Pontoise and CNRS France.}

\date{ }

\begin{document}
\maketitle

\begin{abstract}

We consider the propagation of sine-Gordon kinks in a planar curved
strip as a model of nonlinear wave propagation in
curved wave guides. The homogeneous Neumann transverse boundary
conditions, in the curvilinear coordinates, allow to assume
a homogeneous kink solution. Using a simple collective
variable approach based on the kink coordinate, we show that
curved regions act as potential barriers for the wave and
determine the threshold velocity for the kink to cross.
The analysis is confirmed by numerical solution of the
2D sine-Gordon equation.

\end{abstract}


\bigskip

Recent advances in micro-structuring and nano-structuring technology 
have made it possible to fabricate various low-dimensional systems with 
complicated geometry. Examples are photonic crystals with embedded  
defect structures such as microcavities, wave guides  and wave guide 
bends \cite{souk}; narrow constructions (quantum dots and channels) 
formed at semiconductor heterostructures \cite{reed}, 
magnetic nanodisks, dots  and rings \cite{shinjo,klaeui}, {\it etc}.  

It is well known that the wave equation  subject to
Dirichlet boundary conditions  has  bound states  in straight channels 
of variable width \cite{schult} and in curved channels of constant 
cross-section \cite{gold}. Spectral and transport characteristics of 
quantum electron channels \cite{vakh} and  wave guides in photonic 
crystal \cite{mekis} are essentially modified by the existence of 
segments with finite curvature.
   
Until recently there have been a few  theoretical and numerical studies 
of the effect of curvature on properties of nonlinear excitations. 
The dynamics of a ring shaped Josephson fluxons and their collisions  
was studied in \cite{CO5,CO3,CL}. Nonlinear whispering gallery modes
for a nonlinear Maxwell equation in microdisks were investigated in  
\cite{harayama}, the excitation of whispering-gallery-type electromagnetic 
modes by a moving fluxon in an annular Josephson junction was found in 
\cite{ustinov}. Nonlinear localized modes in two-dimensional photonic 
crystal wave guides were studied in \cite{mingkivsh}. A curved chain of 
nonlinear oscillators  was considered in \cite{curve} and it was shown 
that the interplay of curvature and nonlinearity leads to a symmetry 
breaking when an asymmetric stationary state becomes energetically more 
favorable than a symmetric stationary state. Propagation of Bose-Einstein 
condensates in magnetic wave guides was found quite recently in 
\cite{leanhardt}. Single-mode propagation was observed along homogeneous 
segments of the wave guide while geometric deformations of the 
microfabricated wires lead to strong transverse excitations.
  
The aim of this  article is to study the motion of sine-Gordon (sG)  
solitons moving in a two dimensional finite domain. Specifically we 
treat a planar curved wave guide whose width is much smaller than its 
entire length. We consider homogeneous Neumann boundary 
conditions (zero normal derivative) on the boundaries of the domain.
Using a simple collective variable analysis based on the kink position
we show that a region of non zero curvature in a wave guide induces a 
potential barrier for the wave as it can be verified in the 2D simulations
made by Femlab finite element software. This is different from the case of 
transverse Dirichlet boundary conditions where studies on the
(linear) Schr\"odinger equation show the existence of a localized mode
which will trap waves in the curved region \cite{gold}. 

In the following we study the sine Gordon equation 
as a model for nonlinear wave propagation in planar curved 
wave guides. 
A physically relevant example is a Josephson junction constructed
of two straight segments joined by a bent section (see Fig. 1).
In two spatial dimensions and disregarding the effects of loss and 
external power inputs the sG equation reads
\begin{equation}
  \label{eq:sG2D}
  \frac{\partial^{2} \phi}{\partial t^{2}} - 
  \frac{\partial^{2} \phi}{\partial x^{2}}
  - \frac{\partial^{2} \phi}{\partial y^{2}} + \sin(\phi) = 0 ,
\end{equation}
together with boundary conditions and two initial conditions. For a 
Josephson junction the function $\phi(x,y,t)$ is the normalized phase 
difference of the Cooper pair wave functions across the insulating barrier. 

\begin{figure}[ht]
    \centering
    \includegraphics[width=8cm,height=2.8cm]{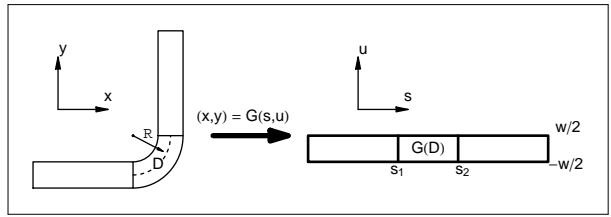}
    \caption{The bend space for the 2+1 D sG equation.
    \label{fig:bend}}
\end{figure}

To solve equation (\ref{eq:sG2D}) in the special domain shown in Fig.
\ref{fig:bend}, we transform the spatial coordinates $(x,y)$ into $(s,u)$, 
where $s$ is arc length of the central line and $u$ is the transversal 
coordinate orthogonal to $s$. A simple geometry with a curved section, $D$, 
of constant radius $R$ has been considered here, where the curvature is 
given by 
\begin{equation}
\label{eq:curvature}
\gamma(s) = \left\{ 
\begin{tabular}{cll}
$\displaystyle -\frac{1}{R},$ & \quad \textrm{when} & $s_1 \leq s \leq s_2.$ \\
$0,$ & \quad \textrm{elsewhere} & \\
\end{tabular} 
\right.
\end{equation}

We parameterize the centerline according to $(x,y)=\sigma(s)=[a(s),b(s)]$ 
and choose the normalized parameterization such that $\|\sigma'(s)\|=1$. 
The unit normal vector $\mathbf{n}$ then becomes $[-b'(s),a'(s)]$. 
Considering for simplicity wave guides with the width smaller than 
the radius of the curvature $R>w/2$, we can describe the wave guide 
domain by means of the following set of coordinates 
$(x,y)=[a(s) - u b'(s),b(s) + u a'(s)]$ as in Ref. \cite{gold}. 

By $\ell$ we denote the total length of the strip and $0 \le s \le \ell$.
The curved region $D$ corresponds to the interval $s_1 \le s \le s_2$.
The width of the strip is $w$ so that $-w/2 \le u \le w/2$. We can determine a 
function $\theta(s)$ such that the tangent vector in the curved region can 
be written in the form $\mathbf{t}$=$\{\cos[\theta(s)],\sin[\theta(s)]\}$,
where $\theta(s)=(\pi/2)(s-s_1)/(s_2-s_1)$. The signed curvature $\gamma(s)$ 
now becomes $\gamma(s) = a'(s) b''(s) - a''(s) b'(s) = \theta'(s)$.
The solution space $[0;\ell] \times [-w/2;w/2]$ we denote by  $\Omega$.
It is Riemannian with the metric $g_{i,j}$, where $i$ and $j$ are $s$ or $u$. 
The determinant of the Jacobian matrix of the variable change leads to
$g=1-u\gamma(s)$.

The Lagrange function of the sine Gordon equation in
Cartesian coordinates $(x,y)$ can be transformed into the new 
variables $(s,u)$ 
\begin{eqnarray}
    L\{\phi\}= \int_{0}^{\ell} \int_{-\frac{w}{2}}^{\frac{w}{2}} \bigg\{ 
    \frac{1}{2} \phi_{t}^2 -  \frac{1}{2 g^2} \phi_{s}^2 - 
    \frac{1}{2} \phi_{u}^2 -  \nonumber \\
    \bigg. [1-\cos(\phi)] \bigg\} g du ds \; .
    \label{eq:Lagrange}
\end{eqnarray}

\noindent{Variation of the above Lagrange function provides the 
sG equation for $\phi$ in the coordinate space $(s,u,t)$},
\begin{equation}
    \frac{\partial^2 \phi}{\partial t^2} - 
    \frac{1}{g} \frac{\partial}{\partial s} \left(\frac{1}{g} \frac{\partial 
    \phi}{\partial s}\right) -
    \frac{1}{g} \frac{\partial}{\partial u} \left(g \frac{\partial 
    \phi}{\partial u}\right) + \sin(\phi) = 0. 
    \label{eq:sGcurved}
\end{equation}
\noindent{To this we add the Neumann boundary conditions corresponding to  
a Josephson junction with no magnetic field or external current 
\begin{equation}
\frac{\partial \phi}{\partial u} = 0 , \qquad
\mbox{on the boundary} ~ \partial \Omega . 
\label{eq:BC}
\end{equation}

Both eqs. (\ref{eq:sG2D}) and (\ref{eq:sGcurved}) can be solved 
numerically in their respective domains. However first 
we investigate kink propagation for the sG equation (\ref{eq:sGcurved}) in the 
curvilinear coordinate space using a simple collective coordinate 
approach for the kink position. 
We assume that our solution $\phi$ only depends on $s$ and $t$, neglecting
the variation along the $u$-direction, $\phi_u = 0$. We denote this 
solution as $\phi_{0}$=$\phi_{0}(s,t)$ and it reads 
\begin{equation}
    \phi_{0}(s,t) = 4 \mbox{arctan} ~e^{s - S(t)},
    \label{eq:phi0}
\end{equation}
\noindent{where $S(t)$ is the position of the center part of the wave.}

The evolution of $S$ is given by the reduced Lagrangian obtained by
substituting $\phi_{0}$ in eq. (\ref{eq:phi0}) and 
performing the spatial integrations over $s$ and $u$. We have
$L = L\{\phi_{0}\}$. 
The reduced Lagrangian is then
\begin{equation}\label{eq:lagr}
L = 4 w  \left(\frac{d S}{dt}\right)^2 - U(S),
\end{equation}
where the effective potential energy $U(S)$ is
\begin{eqnarray}
&&U = \int_{0}^{\ell} \int_{-\frac{w}{2}}^{\frac{w}{2}} 
\left[ \frac{1}{2 g^2} \phi_{s}^2 + [1-\cos(\phi)] \right] g du ds = \nonumber \\
&&4w + 2R\log \frac{1+w/(2R)}{1-w/(2R)} \int_{s_1}^{s_2} \mathrm{sech}^2 (s-S) ds \approx \nonumber \\
&&4w + \frac{w^3}{6R^2} ~ \frac{\sinh (s_2-s_1)}{\cosh (S-s_1) \cosh (s_2-S)}.
\end{eqnarray}
From this expression it is clear that the term $U$ and therefore the 
curved section acts as a potential barrier for the kinks. The maximum of
the barrier is $w^3/3R^2$.

\begin{figure}[h]
\centering
\begin{tabular}{c}
\includegraphics[width=5.cm,height=4.2cm,angle=0]{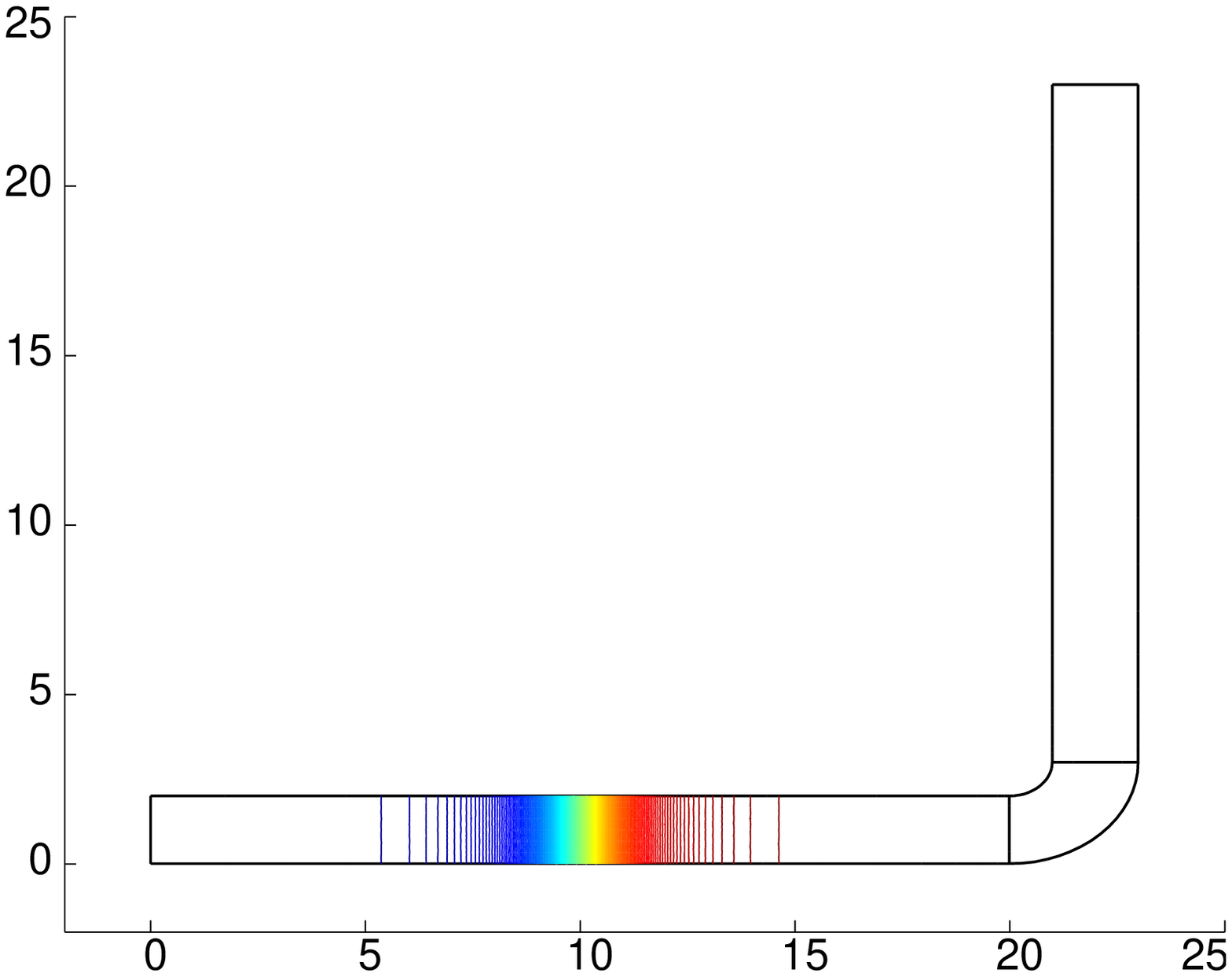}
 \\
\includegraphics[width=5.cm,height=4.2cm,angle=0]{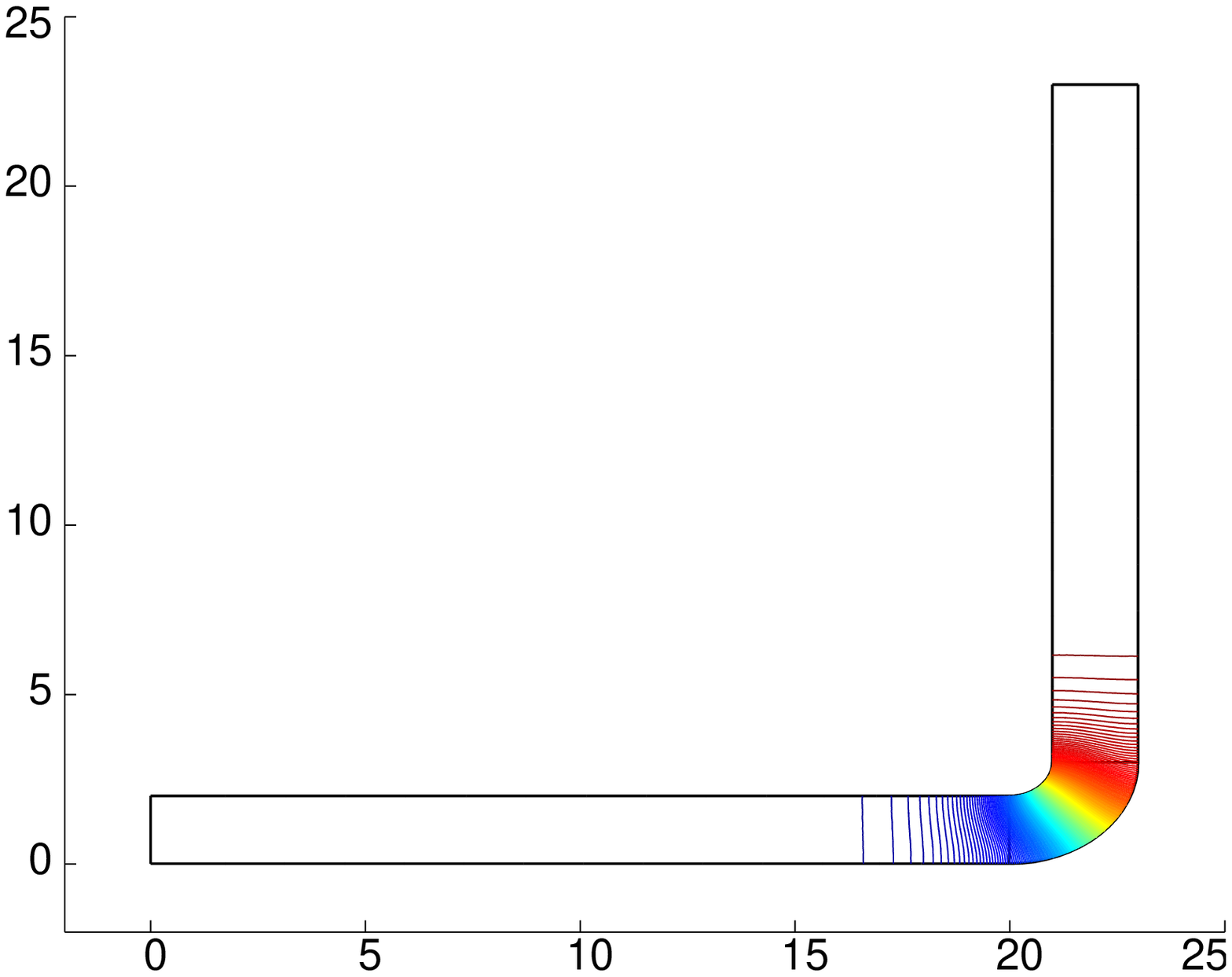}
 \\
\includegraphics[width=5.cm,height=4.2cm,angle=0]{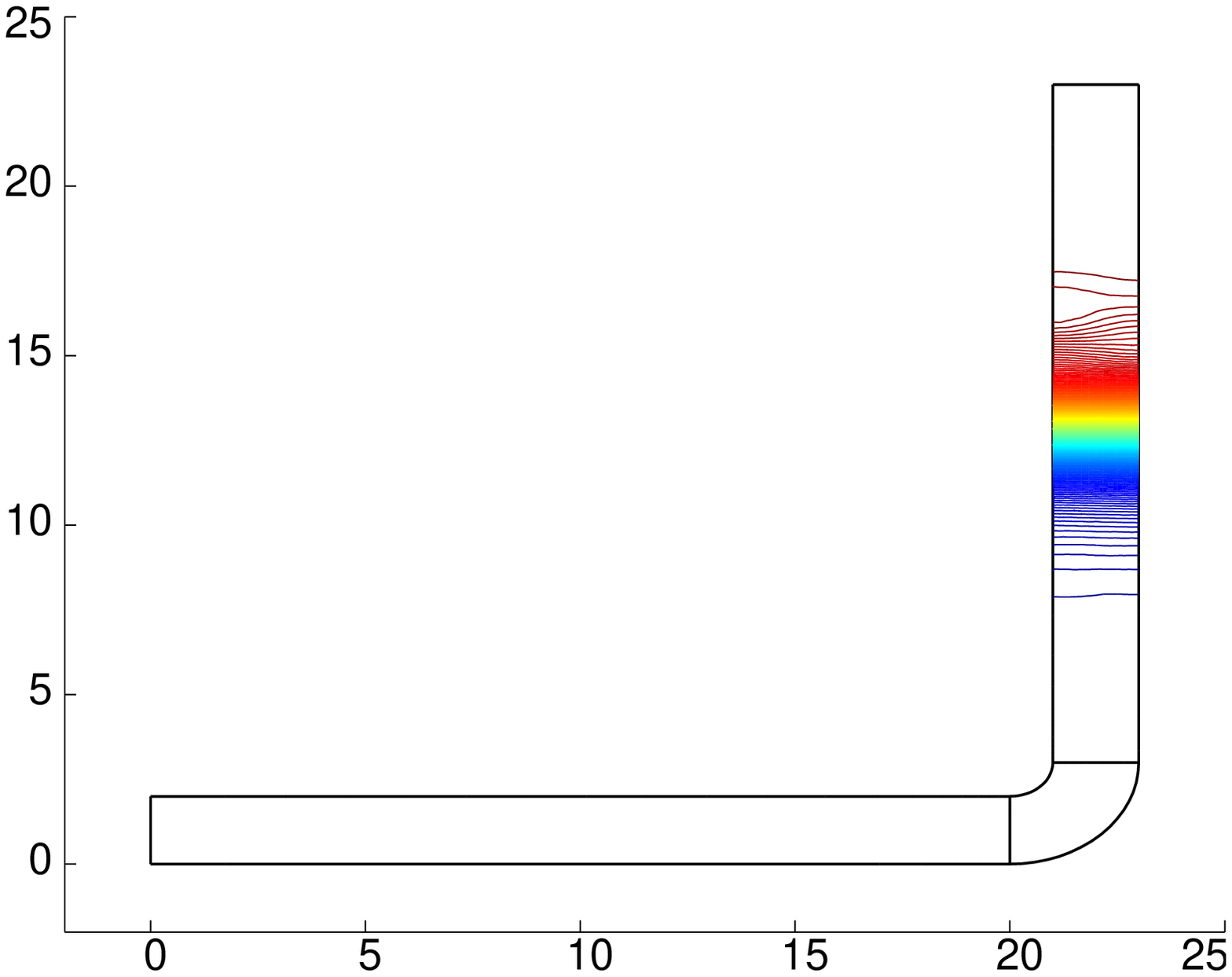}
\end{tabular}
\caption{The wave is transmited for $R=2$, $w=2$ and
$v_0=0.25$, while $v_c\approx 0.29$.\label{stack1}} 
\end{figure}

The initial kinetic energy of the kink, which is located in the 
straight section, far away from the curved region and from the 
domain edge to avoid interference from the boundaries, with initial 
velocity $v_0$, is 
\begin{equation}
\label{eq:kinetic}
T_0 = \int_0^{\ell} \int_{-w/2}^{w/2} \frac{1}{2} \phi_{t}^2~du~ds
= 4wv_0^2.
\end{equation}
When $T_0$ is large enough to overcome the potential 
$U$ the wave crosses the barrier running along the channel 
as shown in the contour-plot sequence of Fig. \ref{stack1}. 
The minimal velocity 
for the kink to cross the barrier is calculated by comparing the 
kinetic energy $T_0$ to the potential energy $U$. It is approximately
\begin{equation}
v_c \approx \frac{w }{\sqrt{12}R}\sqrt{\mathrm{tanh} \frac{s_2-s_1}{2}}
\end{equation}

On the other hand when the wave is not initially 
fast enough, the potential barrier reflects the
wave and it returns with a negative velocity 
(see Fig. \ref{stack2}).

\begin{figure}[h]
\centering
\begin{tabular}{c}
\includegraphics[width=5.cm,height=4.2cm,angle=0]{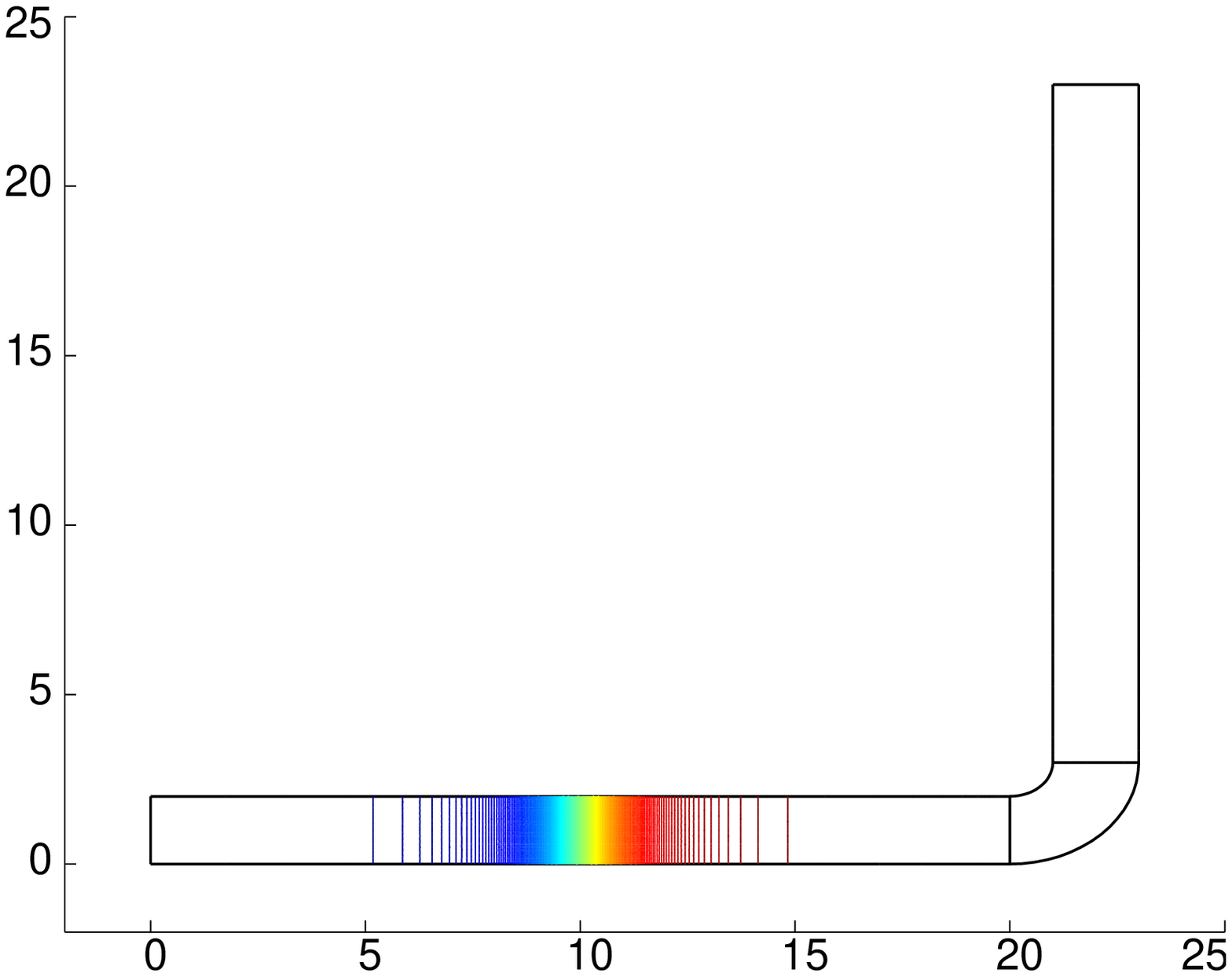}
 \\
\includegraphics[width=5.cm,height=4.2cm,angle=0]{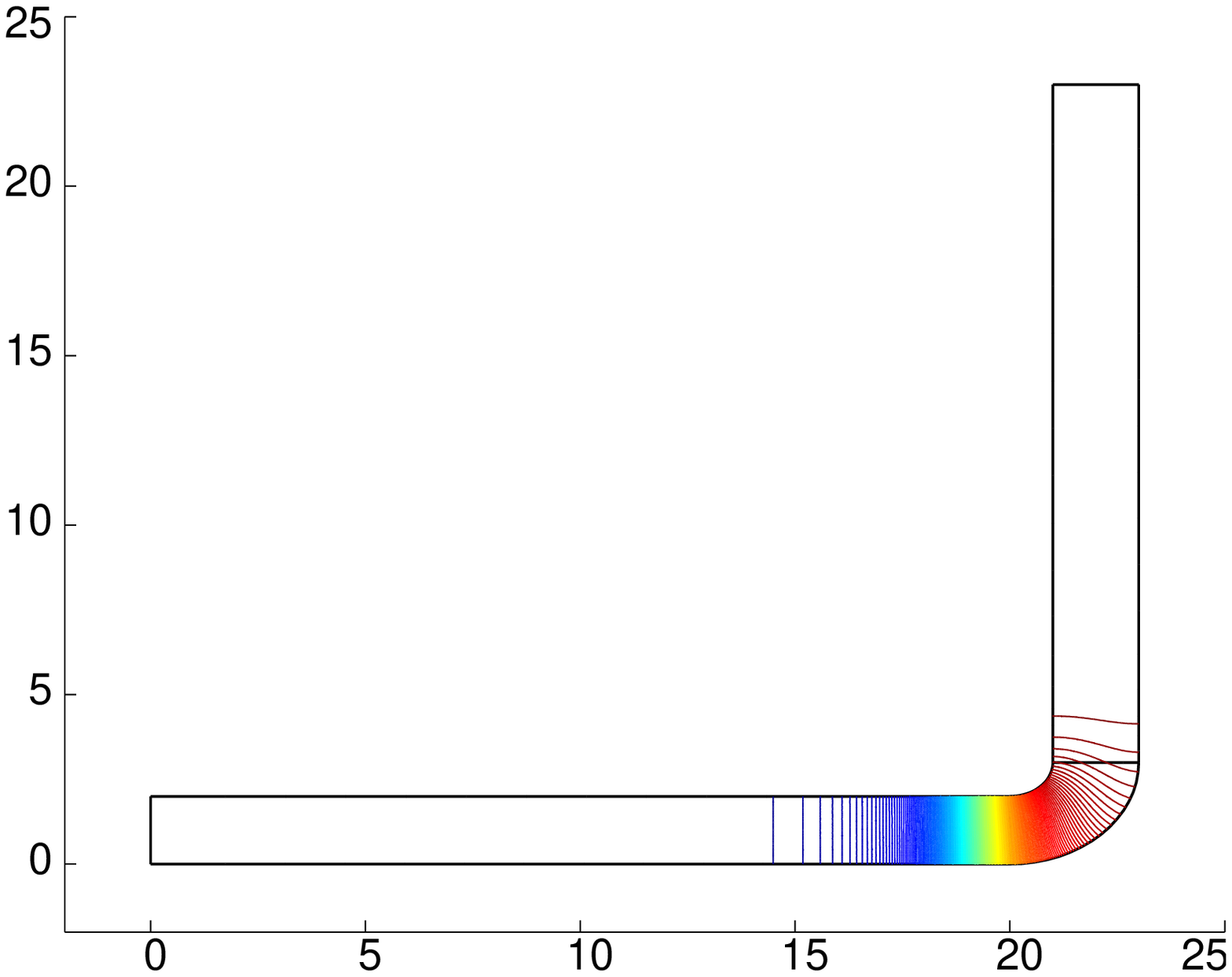}
 \\
\includegraphics[width=5.cm,height=4.2cm,angle=0]{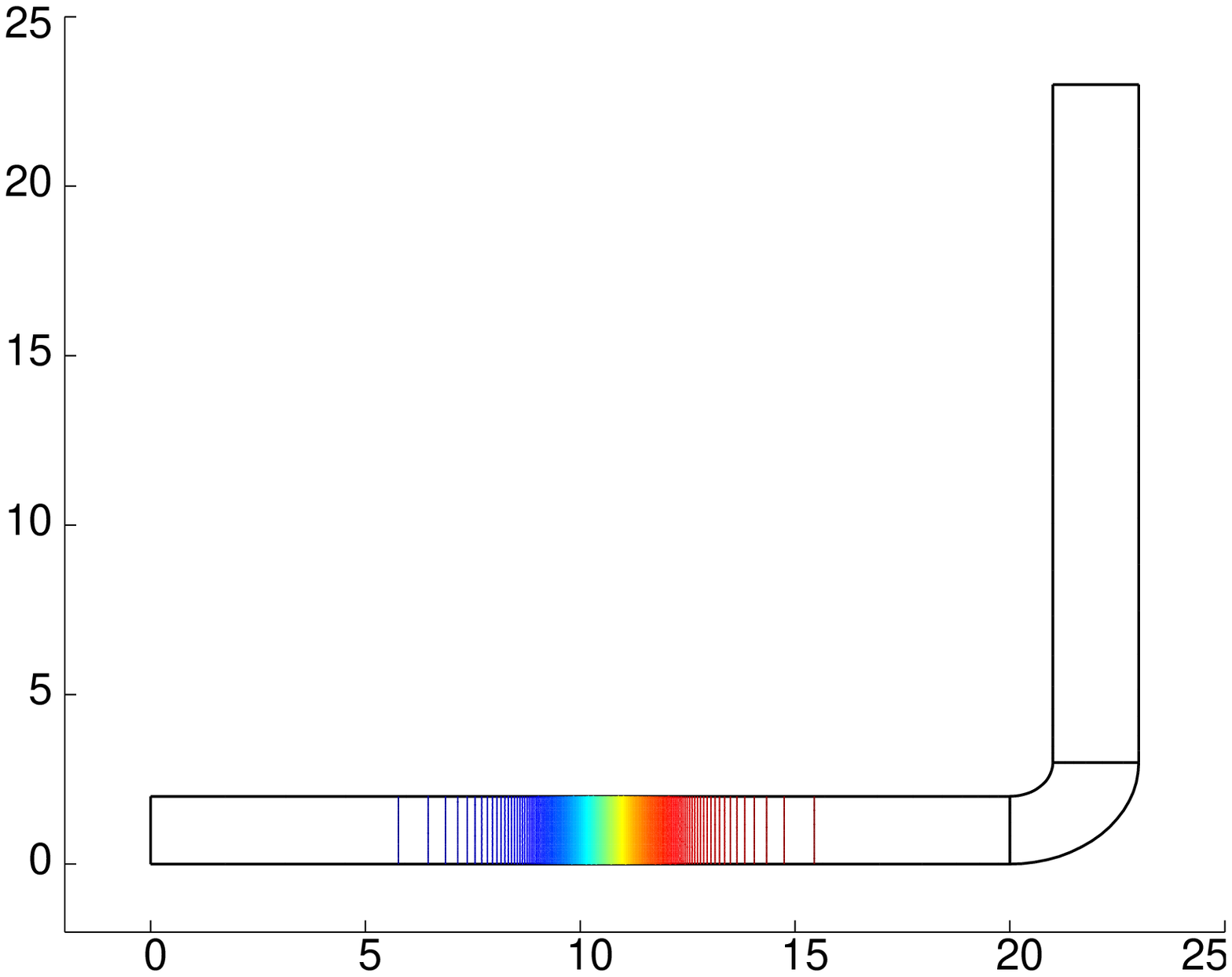}
\end{tabular}
\caption{The wave is reflected for $R=2$, $w=2$ and
$v_0=0.25$, while $v_c\approx 0.29$.\label{stack2}} 
\end{figure}

We now proceed to compare quantitatively the predictions
of this simple theory with the solution of the 2D problem.
We do this by plotting the phase space $(S,dS/dt)$ both for
the collective variable dynamics (energy levels) 
and the kink position $S$ and velocity $dS/dt$ estimated 
from the 2D solution. 
The initial position of the kink is fixed well away from
the curved region $20 <S<20+R\pi/2$. Then the initial velocity 
of the soliton determines completely its trajectory as can be 
seen in the phase space $(S,dS/dt)$ shown in Fig. \ref{phase}. 
When the initial velocity is lower than $v_0 \approx 0.2$ the orbits do not 
cross the curved region. On the other hand for higher initial 
velocities the orbits span the complete circuit through the $S$ 
axis. Due to the existence of effective potential barrier the 
velocity of kink decreases in the curved segment. Fig. \ref{phase} 
shows that the orbits calculated from the numerical solution of 
the 2D system are are in very good agreement with the
ones given by the collective coordinate approach.

\begin{figure}[h]
\centering
\includegraphics[width=7.cm,height=5.cm,angle=0]{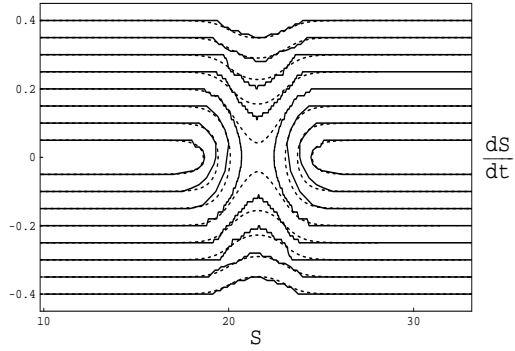}
\caption{Phase space $(S,dS/dt)$ of the kink position and velocity for
$R=2$ and $w=2$. Comparison of simulations (continuous lines)
and the collective coordinates (dashed lines). \label{phase}}
\end{figure}

The simple analytical approach developed above provides a good 
description of the kink dynamics for small velocities, otherwise
the ansatz used in Eq. (\ref{eq:phi0}) may be too crude for large 
ones. In that case the kink ansatz should also depend on the variable 
$u$ as the contour plot of Fig. \ref{contour} shows.
The distance in $s$ covered by the wave increases as
$u$ decreases from $w/2$ to $-w/2$. Thus the soliton velocity depends 
on the level at $u$ axis and Lorenz contraction induces the different kink 
widths observed for different values of $u$.

\begin{figure}[h]
\centering
\includegraphics[width=5.cm,height=4.2cm,angle=0]{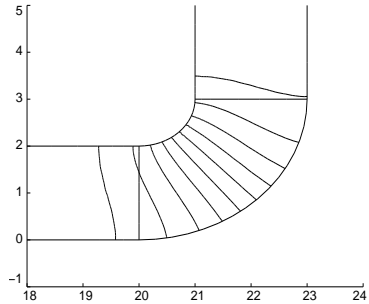}
\caption{Contour plot of the wave at the curved section for $R=2$, 
$w=2$ and $v_0=0.20$.\label{contour}} 
\end{figure}

The numerical simulations are done in the cartesian coordinates
using the Femlab finite element software \cite{femlab} to define the 
domain $\Omega$ and discretizing it by a triangular mesh. The time 
evolution has been performed with a differential equation solver. 

We consider a nonlinear wave propagation in a curved planar wave 
guide using as a model kink solutions for the sine-Gordon equation.
Specifically the domain considered $\Omega$ in the plane 
is made of two rectangular regions joined by a bent section of
constant curvature. The transverse homogeneous Neuman boundary 
conditions allow us to consider a homogeneous kink solution. 
Following this we develop a simple collective variable theory 
based on the kink position. This shows that curved regions act as 
potential barriers for the waves in contrary to the case of Dirichlet 
boundary conditions. We calculate the treshold velocity for the kink
to cross and it is in excellent agreement with the solutions of the
full problem calculated with Femlab finite element program. The
phase-space of the system matches also well the one obtained from
the numerical solution.

This study shows that it is possible to trap fluxons in a Josephson 
junction, in the region between two successive bends.
Choosing a conveniently small damping term, one may decrease the kinetic 
energy of the soliton as it comes into this region so that it will
stay there. This feature could be applied to
electronic devices for storing binary data. 

\medskip

C.G. and Yu.~B. G. acknowledge the hospitality of the Technical University 
of Denmark and of the University of the Basque Country helping with all kind 
of facilities during the investigation period. Financial 
support was provided by the LOCNET Program (HPRN-CT-1999-00163),
a project from the University of the Basque Country (UPV00100.310-E-14806/2002).


\bibliographystyle{unsrt}

\end{document}